%% file: arxiv-erratum-corrected.tex
\documentclass[prd,aps,nofootinbib,preprint]{revtex4-1}
\usepackage{amsmath,amssymb,amsfonts,color,graphicx,graphics,latexsym,placeins,epsfig,multirow}
\usepackage{array}
\newcolumntype{P}[1]{>{\centering\arraybackslash}p{#1}}
\usepackage{footmisc}
\usepackage[compatibility=false]{caption}
\usepackage{subcaption}
\usepackage{comment}
\usepackage{bbold}
\usepackage{enumitem}
\setdescription{leftmargin=8pt}
\usepackage{bbold}
\usepackage{hyperref}
\hypersetup{
	colorlinks=true,       
	linkcolor=blue,          
	citecolor=blue,        
	urlcolor=blue           
}

\usepackage{natbib}
\setcitestyle{square, comma, numbers, sort&compress}
\usepackage{float}
\usepackage{amsmath,amssymb}
\newcommand{\A}{\mathcal{A}}
\newcommand{\E}{\mathcal{E}}

\newcommand{\G}{\mathcal{G}}
\newcommand{\B}{\mathcal{B}}
\newcommand{\ba}{\begin{align}}
\newcommand{\ea}{\end{align}}

\def\3nab{\tilde{\nabla}}

\def\be {\begin{equation}}
\def\ee {\end{equation}}
\def\ba {\begin{eqnarray}}
\def\ea {\end{eqnarray}}

\newcommand{\bra}[1]{\left(#1\right)}

\newcommand{\sfr}[2]
{{\textstyle\frac{#1}{#2}}}

\newcommand{\lc}{\varepsilon}
\newcommand{\lb}{\{}
\newcommand{\rb}{\}}

\renewcommand{\H}{{\mathcal H}}
\newcommand{\barray}{\begin{array}}
\newcommand{\earray}{\end{array}}

\newcommand{\sss}[1][0.035cm]{\hspace*{#1}}
\newcommand{\udot}{{\mathcal A}}
\newcommand{\bea}{\begin{eqnarray}}
\newcommand{\eea}{\end{eqnarray}}
\newcommand{\f}[2]{\textstyle\frac{#1}{#2}}

\newcommand{\udota}{{\cal A}}

\usepackage[dvipsnames]{xcolor}

\begin{document}
\title{\Large \bf Prospects for cosmological constraints using gravitational wave memory}

\author{Indranil Chakraborty}
\email{indranil.phy@iitb.ac.in}
\thanks{Equal contribution to this work.}

\author{Susmita Jana}
\email{susmitajana@iitb.ac.in}
\thanks{Equal contribution to this work.}

\author{S. Shankaranarayanan}
\email{shanki@iitb.ac.in}

\affiliation{Department of Physics,  Indian Institute of Technology Bombay, Mumbai 400076, India}
\begin{abstract}
{The $\Lambda$CDM model has long served as a robust and predictive framework for cosmology, successfully explaining a wide range of observations, including the accelerated expansion of the Universe. However, discrepancies in cosmological parameter estimates and recent findings, such as those from DESI, hint at potential deviations from $\Lambda$CDM. Gravitational wave (GW) observations offer an independent method to probe the nature of dark energy, leveraging GWs from compact binary mergers as standard candles.
In this study, we demonstrate that the integrated GW memory over cosmological distances encodes a unique imprint of the background spacetime. Unlike previous analyses, our approach captures non-linear dependencies on cosmological quantities, resulting in an enhancement of the integrated GW memory by a factor of 100 for high-redshift sources—well within the sensitivity range of next-generation detectors like Cosmic Explorer and the Einstein Telescope.
We find that despite the diminishing strength of individual GWs at high redshifts, their cumulative effect leads to a significant amplification, akin to the integrated Sachs-Wolfe effect, offering a potential new avenue for cosmological studies. By examining a range of dark energy models, we reveal that GW memory is potentially highly sensitive to the underlying cosmological framework, making it a promising probe of dark energy. This novel approach presents the possibility of a fresh perspective to address persistent cosmological tensions, and the nature of dark energy.}
\end{abstract}
\maketitle



\section{Introduction}
\label{sec:intro}
The $\Lambda$CDM model has become the cornerstone of modern cosmology due to its impressive ability to shed light on the structure and evolution of the Universe~\cite{2008-Weinberg-Book,2013-Peter.Uzan-Book,Rees:2022znv}. However, it also presents several unanswered questions~\cite{2022-Melia-Review}. $\Lambda$CDM model addresses the flatness of the Universe and the uniformity of the cosmic microwave background (CMB) radiation by incorporating 
inflation~\cite{Ellis:2023wic}. However, the specifics of inflation
remain unclear~\cite{Martin:2013tda,Odintsov:2023weg}. The second puzzle is dark matter --- a particle inferred solely from interaction with gravity~\cite{Profumo:2019ujg,Rajendran:2022kcs,Carr:2020xqk}. 
Third, the current acceleration of the universe~\cite{Sahni:1999, Peebles:2002gy,Padmanabhan:2002ji,Joyce:2016vqv,Motta:2021hvl,Weinberg:1988cp,Weinberg:2000yb,Carroll:2000fy,DiValentino:2021izs,Shah:2021onj,Schoneberg:2021qvd,Kamionkowski:2022pkx,2022-Shanki.Joseph-GRG,2012-Lima.etal,2016-Moresco.etal-JCAP}. Recent DESI results hint at slight modifications from $\Lambda$CDM~\cite{DESI:2024_I,DESI:2024l-II}.

The direct detection of gravitational waves (GWs) opened an unprecedented channel to probe some of these open questions in cosmology~\cite{2009-Sathyaprakash.Schutz-LRR}. For instance, the first GW detection from merging black holes (BHs) has reignited the possibility that primordial BHs (PBHs) may constitute most of the dark matter~\cite{2016-Sasaki.etal-PRL,2016-Bird.etal-PRL,2016-Hayasaki.etal-PASJ}. The first binary neutron star (BNS) event (GW170817)~\cite{LIGOScientific:2017zic,LIGOScientific:2017vwq} has provided a new probe of Hubble constant ($H_0$)~\cite{Schutz:1986gp,LIGOScientific:2017adf,2017-Chen.Holz-Nature}. It is expected that GW observations alone can measure $H_0$ with $1\%$ accuracy~\cite{DelPozzo:2011vcw,Nissanke:2013fka}. 

GW observations can provide an independent way to understand the nature of dark energy as the GW signal from the merging binary systems contains direct information about the luminosity distance to the source~\cite{Krolak:1987ofj,2009-Sathyaprakash.Schutz-LRR}. Most of the analysis has focused on the redshifting of the GW frequencies and dilution of the GW amplitude. However, as GWs propagate through cosmological distances, they induce subsequent GWs, creating successive GWs referred to as \emph{GW memory}~\cite{1985-Braginsky-JETP,2010-Favata-CQG}. In this work, we show that GW memory can be used as a probe to distinguish between cosmological models, building on the growing relevance of GWs in cosmology.

GW memory refers to a non-vanishing term in the
perturbation, leaving an imprint of the passage of the GW~\cite{1985-Braginsky-JETP,2010-Favata-CQG}. In asymptotic Minkowski space-time, this corresponds to
\begin{align}
\Delta h_{+, \times}^{\mathrm{mem}}=\lim _{t \rightarrow+\infty} h_{+, \times}(t)-\lim _{t \rightarrow-\infty} h_{+, \times}(t),
\end{align} 
where, $h_{+},\, h_{\times}$ is the plus and cross mode of the GW. This effect manifests as a permanent change in separation (displacement memory), or change in relative velocity (velocity memory), between free test masses \cite{1974-Zeldovich-JETP,1991-Christodoulou-PRL,1987-ThorneBraginskii-Nature,2010-Favata-CQG,2009-Favata-ApJL,2017-Zhang.etal-PRD}. This effect has been extensively studied in asymptotically flat (AF) spacetimes owing to its relation to Bondi-Metzner-Sachs (BMS) symmetries~\cite{2014-Strominger.Zhiboedov-JHEP,Flanagan:2015,Bhattacharjee:2019jaf}. Comprising of the low-frequency component of the emitted GWs, gravitational memory is  determined by solving the sourced Einstein's field equations~\cite{Flanagan:2015, Nichols:2017, Mitman:2020,Strominger_lectures:2017,Compere:2018,Thorne:1992, Wiseman:1991,Jenkins:2021, Mukhopadhyay:2021, 2022-Jokela.Sarkkinen-PRD,Shore:2018,Siddhant:2020,Chakraborty:2021,Heisenberg_memory:2023,Donnay:2018,Boersma:2020,Zhang:2017,Zhang:2018,Hait:2022,Bhattacharya:2023-SN}.
While BMS symmetries and memory in AF spacetimes provide important insights into infrared physics, they can not be directly applied to cosmological settings, as one needs to take into account the  background curvature. 
Previous studies on memory employing perturbative approaches have shown the enhancement attributable to the redshift factor~\cite{Tolish_Wald_cosmology:2016, Bieri_cosmology:2017} along with the presence of tail terms \cite{Chu:2016, Kehagias:2016,2022-Jokela.Sarkkinen-PRD}. 

This leads to the following questions: Is there a unified treatment of GW memory for a class of spacetimes? What role does curvature play? 
Can the cumulative effect of the successive GWs generated in the cosmological space-times lead to observable signatures? Interestingly, in Ref.~\cite{2023-Jana.Shanki-PRD} the authors obtained a master equation 
for electromagnetic (EM) memory in generic spacetimes, including Kerr. Since gravity is non-linear, extending the analysis to general relativity is not straightforward. However, as we show in this work, it is possible to obtain a \emph{master equation} (ME) for GW memory for Locally Rotationally Symmetric (LRS) of type II spacetimes~\cite{1996-vanElst.Ellis-CQG}. 

A space-time is considered to be LRS if, at each point, a one-parameter group of rotations preserves the Riemann tensor and its derivatives up to third order~\cite{1996-vanElst.Ellis-CQG}. LRS-II 
is characterized by the fact that the magnetic Weyl tensor $H_{a b}$, vorticity ($\omega_{a b}$), and 2-sheet twisting $\xi$ all vanish. 
These spacetimes are both time and space-dependent and contain
many physically interesting solutions like spherically-symmetric perfect fluids, Bianchi I, III and Lemaitre-Tolman-Bondi (LTB) cosmologies, Kantowski-Sachs, and the flat and hyperbolic Friedmann-Lema$\hat{i}$tre-Robertson-Walker (FLRW) models~\cite{Ellis_JMP:1967,1996-vanElst.Ellis-CQG}. Using the semi-tetrad approach~\cite{2004-Clarkson.P.K.S-APJ,2003-Clarkson.Barrett-CQG,clarksonlrs,2019-Goswami.Ellis-CQG}, we obtain GW memory for LRS-II spacetimes. For FLRW, 
we demonstrate that their cumulative effect over cosmological time leads to an enhancement similar to the integrated Sachs-Wolfe effect in CMB~\cite{2008-Weinberg-Book}. The enhancement can help constrain cosmological models with Cosmic Explorer (CE)and Einstein Telescope (ET)~\cite{Branchesi:2023,Evans:2023-CE}. \smallskip

We use semi-tetrad formalism to derive ME for GW memory~\cite{2004-Clarkson.P.K.S-APJ,2003-Clarkson.Barrett-CQG,clarksonlrs,2019-Goswami.Ellis-CQG}. An advantage of this approach compared to traditional perturbation theory is that it avoids switching between gauges when evolving metric perturbations along the comoving observer~\cite{2012-ellis_maartens_maccallum-Book}. This is due to the Stewart-Walker Lemma~\cite{SW}, which states that any geometrical or thermodynamic quantities that vanish in the background are inherently gauge-invariant. Hence, the GW memory we evaluate is \emph{gauge invariant}.

In this formalism, $4-$D spacetime is decomposed using the fluid 4-velocity $(u^{a})$ and a preferred spatial direction $(n^{a})$ in the 3-space. $u^a$ and $n^a$ satisfy the conditions $u^{a}u_{a}=-1, n^{a}n_{a}=1, u^a n_a=0$. [We use $(-,+,+,+)$  signature  and set $8 \pi G = c= 1$.] The projection tensor:
\begin{equation}
N_{ab} = g_{ab} + u_{a}u_{b} - n_{a}n_{b}  \, ,
\label{eq:1+1+2}
\end{equation} 
projects vectors orthogonal to $u^a$ and $n^a$ onto 2-D surface~\cite{2019-Goswami.Ellis-CQG}. In this formalism, any quantity can be split into scalars, 2-vectors, and projected, symmetric, and trace-free 2-tensors, where the latter two components are defined on the 2-D surface.
Specifically, the $4-$D spacetime is described by the kinematic variables --- expansion, shear, and vorticity associated with $u_{a}$ and $n_{a}$, and components of the Weyl tensor ($C_{abcd}$)~\cite{2004-Clarkson.P.K.S-APJ,2003-Clarkson.Barrett-CQG,clarksonlrs,2019-Goswami.Ellis-CQG}. The electric ($E_{ab}$) and magnetic part ($B_{ab}$) of the Weyl tensor along the 2-D surface carry information about the passing GWs~\cite{2011-Owen.Thorne-PRL,Thorne2}. This provides a natural representation for the GWs where the two traceless-transverse DOF lie on the 2-D surface~\cite{2004-Clarkson.P.K.S-APJ,2003-Clarkson.Barrett-CQG,clarksonlrs,2019-Goswami.Ellis-CQG}.

For LRS spacetimes with spherical symmetry, we can choose $n_{a}$ such that all the geometrical variables on the $2$-D surface $N_{ab}$ vanish~\cite{1996-vanElst.Ellis-CQG}. For the LRS-II spacetimes, which is the focus of this work, the vorticity associated with $u^a$ and $n^a$ also vanish~\cite{1996-vanElst.Ellis-CQG}. Due to the vanishing of the vorticity in LRS-II spacetime, $E_{ab}$ and $B_{ab}$ can be orthogonal, and $E_{ab}$ can be expressed in terms of $B_{ab}$~\cite{2005-Herrera-J.Math.Phys}. As we show, this property helps us to derive the ME for GW memory in these spacetimes. 
\begin{figure}[!t]
\centering
\includegraphics[width=0.85\textwidth]{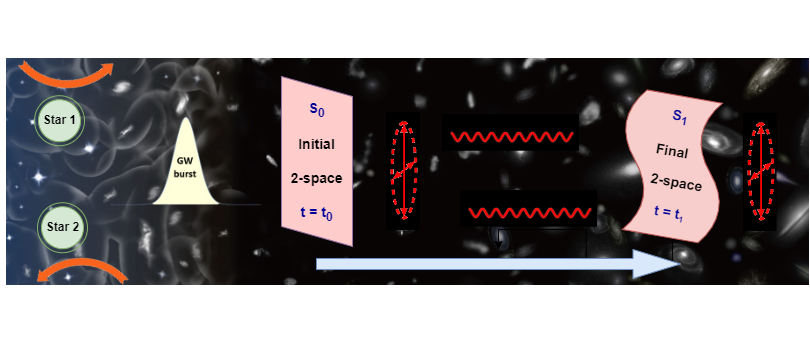}
\caption{\label{fig:FLRW_mem_model} Schematic depiction of GW memory in LRS-II spacetime. We model the generated GWs as a burst occurring at a particular epoch ($t=t_0$). This alters the shear of the 2-space at $t_1 > t_0$. The blue arrow at the bottom depicts the direction of time ($u^a$). The orange arrows denote the trajectory of the stars.}
\end{figure} 

The setup in the LRS-II spacetime can be seen in Fig.~\eqref{fig:FLRW_mem_model}. Consider an LRS-II spacetime where an event generating GWs occurs at $t = t_0$. As the GWs propagate from $t = t_0$ to a later time $t_1 > t_0$, the primary GWs generate successive GWs. This process is captured on the 2-D surface \(N_{ab}\), which is orthogonal to the direction of propagation. We aim to evaluate the change in \(N_{ab}\) for a comoving observer.

To compute $\Delta N_{ab} =N_{ab}(t_1)-N_{ab} (t_0)$, we proceed as follows: Prior to the event ($t < t_0$), the spacetime is purely LRS-II. At $t = t_0$, the event generates GWs, and because our focus is on the successive GWs, the amplitude of the primary GW event serves as the source. Given that the amplitude of the successive GWs is small, we assume that the spacetime is a perturbed LRS-II~\footnote{This has to be contrasted with the covariant perturbation theory in cosmological spacetimes where the growth of tensor and scalar perturbations are generated due to the quantum fluctuations~\cite{Bruni:1992dg,2012-ellis_maartens_maccallum-Book}.}. The GW memory of the perturbed LRS-II spacetime is related to the change in the geometrical quantities associated with the 2-D surface w.r.t the timelike direction $u^a$: $\mathcal{X}_{ab} \equiv \mathcal{L}_{u} N_{ab}$. Since the GWs are symmetric and traceless, we have: 
\begin{align} \label{eq:shear_to_memory}
    \mathcal{X}_{\{ab\}} = 2 \,\Sigma_{ab} \, ,
\end{align}
where $\Sigma_{ab}$ is 2-D shear tensor~\cite{clarksonlrs} and we refer to as the \emph{instantaneous memory tensor} for the comoving (Lagrangian) observers~\cite{Ellis_JMP:1967,1996-vanElst.Ellis-CQG}. Since these are natural observers in cosmology, the semi-tetrad formalism allows us to evaluate the \emph{cumulative memory} \((\Delta N_{ab})\) along $u^a$\cite{2004-Clarkson.P.K.S-APJ,2003-Clarkson.Barrett-CQG,clarksonlrs,2019-Goswami.Ellis-CQG}. 
This identification is crucial for deriving ME for memory in LRS-II. Details in appendix~\eqref{app:memory-shear}.

\section{Geometric quantities}
\label{sec:Geom-quantities}
Before the event ($t < t_0$), the spacetime is described by the following variables~\cite{2003-Clarkson.Barrett-CQG}: 
\be\label{background}
\mathcal{D}_0=\{ \overline{\Theta}, \overline{\phi}, \overline{\Sigma}, \overline{\udot}, \overline{\E}\} \cup \mbox{matter components},
\ee
where, $\overline{\Theta}$ and $\overline{\phi}$ represent expansion along $u^a$ and $n^a$ in LRS-II space-time, respectively, $\overline{\Sigma}$ is the component of shear tensor along $n^a$, $\overline{\A}$ is the acceleration projected along $n^a$ of the comoving observer, and $\overline{\E}$ is electric-Weyl tensor projected  along $n^a$. As mentioned above, all other geometrical variables vanish on the 2-D surface. 

After $t > t_0$, the spacetime is no longer LRS-II. The perturbed LRS-II spacetime is described by the following variables: $\mathcal{D}_{\rm n} \equiv \mathcal{D}_{0} \cup \{\E_a,\, \E_{ab},\, \B_{a},\, \B_{ab} ,\, \Sigma_{ab}\}$, where $\{\E_a,\, \E_{ab},\, \B_{a},\, \B_{ab}\}$ depend on the properties of the primary GWs. $\Sigma_{ab}$ encodes the instantaneous memory due to the primary GWs. The perturbed 2-D surface contains \emph{non-zero components} of $\Sigma_{ab}$ (that carries information about the successive GW), electric and magnetic Weyl $(\E_a,\, \E_{ab},\, \B_{a},\, \B_{ab})$ defined as:
\begin{eqnarray}
    E_{ab}=\E (n_a n_b-N_{ab}/2) +2 n_{(a}\E_{b)} +\E_{ab}, \label{eq:elec-weyl-decom}\\
    B_{ab}=\B (n_a n_b-N_{ab}/2) +2 n_{(a}\B_{b)} +\B_{ab}, \label{eq:mag-weyl-decom}\\
    \sigma_{ab} = \Sigma (n_a n_b-N_{ab}/2) +2 n_{(a}\Sigma_{b)} +\Sigma_{ab} \label{eq:shear-decom}.
\end{eqnarray}
Here, $(\E, \B)$ are scalars, $(\E_a, \B_a)$ are vectors and $(\E_{ab}, \B_{ab})$ are tensors defined on the 2-D surface. The two scalars, vectors and tensors are the 10 components of the Weyl tensor and are not independent. Since, we are interested in GW memory, we will focus only on the tensor perturbations. To do that, we must identify the physical quantities and their dynamical equations. 
 
In semi-tetrad formulation, the perturbed Einstein's equations in LRS-II spacetimes can be written as a set of evolution, propagation and constraint equations of electric $(\E_{a},\, \E_{ab})$ and magnetic parts $(\B_{a},\, \B_{ab})$ of Weyl tensor and instantaneous memory tensor ($\Sigma_{ab}$). The propagation and constraints reduce the number of independent degrees of freedom (DOF). In particular, the propagation equations (along the radial direction, denoted by over-hat, and the 2-space derivatives are given by $\delta$) for $ \E_{a}, \B_{a}$ and their evolution equations, lead to the following constraints:
\begin{align}
\label{eq:EvecTen-HvecTen}
\E_a = \lc_{ab} \, \B^b~;~\E_{ab} = \lc_{ad} \, \B^{d}\,_{b}
\end{align}
where, $\lc_{ab}$ is Levi-Civita tensor in $2-$D surface. Along with the propagation equation for the scalars,
\begin{align}
\hat{\E} = - \delta_{a}\E^{a} -  3 \overline{\phi} \, \overline{\E}/2;~~
\hat{\B} = - \delta_{a}\B^{a} - 3 \overline{\phi} \, \B/2 \, .
\label{eq:EBScalar-hat}
\end{align}
we have only 2 DOF for the tensor perturbations. We are now in a position to obtain the evolution of $\Sigma_{ab}$ for these 2 DOFs. Details in the appendix~\eqref{app:DOF}. 

\section{Master equation for GW memory}
\label{sec:ME-GWmemory}
To arrive at ME, we first use the evolution equation of $\E_{ab}$: 
\begin{eqnarray}
\label{Edot} 
\partial_- \E_{\lb ab\rb} + \partial_-\left[\lc_{c\lb a}\B_{b\rb}^{~~c}\right] 
& = -\lc_{c\lb a}\delta^{c}\B_{b\rb} -\f32 \overline{\E} \, \Sigma_{ab} 
  - \bra{\overline{\Theta} + \f32 \overline{\Sigma}} \E_{ab} \\ 
  &+ \bra{\f12\overline{\phi} + 2 \overline{\udota}}\lc_{c\lb a}\B_{b\rb}^{~~c}  -\f12 (\overline{\mu} + \overline{p}) \Sigma_{ab} \nonumber
\end{eqnarray}
where $``-"$ refers to null coordinate $x_-\equiv t-r$, the matter in the background LRS II spacetime is represented by a perfect fluid with pressure $\overline{p}$ and energy density $\overline{\mu}$. Substituting Eq.~\eqref{eq:EvecTen-HvecTen} in the above equation leads to:
\begin{align}
\label{eq:master1-main}
3 \overline{\E} \, \Sigma_{ab} + \left[2 \overline{\Theta} + 3 \overline{\Sigma} + \overline{\phi} \right]\E_{ab} + \left[\overline{\mu} + \overline{p} \right] \Sigma_{ab} 
= -\G_{ab}.
\end{align}
where, $\G_{ab}\equiv 2\, \epsilon_{c\{a}\delta^c\B_{b\}}$ captures the primary GWs. 

As mentioned earlier, $\Sigma_{ab}$ carries information about the alteration of the 2-D surface. Hence we want to re-write Eq.~\eqref{eq:master1-main} substituting $\E_{ab}$ in terms of $\Sigma_{ab}$. To do so, we use the evolution equation of $\Sigma_{ab}$: 
 \begin{align}
\label{eq:Sigma-dot}
\dot\Sigma_{\lb ab \rb}  
+ \left(2 \overline{\Theta}/3 + \overline{\Sigma}/2\right)\Sigma_{ab} = -\E_{ab} \, ,
\end{align} 
where dot denotes derivative w.r.t timelike direction $u^a$. Substituting Eq.~(\ref{eq:Sigma-dot}) in the constraint Eq~(\ref{eq:master1-main}) leads to the following ME for GW memory: 
\begin{align}
\label{eq:master-main}
\left[2 \overline{\Theta} + 3 \overline{\Sigma} + \overline{\phi}  \right] \left[\dot\Sigma_{\lb ab \rb} +  
\left(\f23  \overline{\Theta} + \f12  \overline{\Sigma} \right) \Sigma_{ab} \right]  
- \left[3  \overline{\E} \, +  \overline{\mu} +  \overline{p} \right] \Sigma_{ab} = \G_{ab} \, .
\end{align}
This is the key expression of this work, regarding which we want to discuss the following points: To begin with, ME delineates how $\G_{ab}$ manifests its presence through $\Sigma_{ab}$. ME relates $\Sigma_{ab}$ with the geometrical quantities of the background spacetime and the primary GW. When $\G_{ab} = 0$, $\Sigma_{ab}$ does not evolve. This is because, without the primary GW, spacetime remains LRS at all times. Thus, ME resembles a forced system, where the primary GW provides the forcing term. Such forced systems have been discussed earlier~\cite{1985-Braginsky-JETP,Siddhant:2020}. 

In addition, ME is valid for all LRS-II spacetimes, including Minkowski, FLRW, and LTB cosmologies \footnote{In deriving the above expression, we have assumed acceleration ($\A$) vanishes. In appendix~\eqref{app:formulation-masterEq}, we have derived the expression for nonzero acceleration ($\A$).}. Earlier works on GW memory predominantly focused on AF spacetimes described by the BMS metric. This metric framework is well-suited for studying GW memory effects for isolated sources, where the spacetime approaches flatness at large distances from the source~\cite{2013-Bieri.Garfinkle-PRD,2014-Tolish.Wald-PRD,2014-Strominger.Zhiboedov-JHEP}. The semi-tetrad formulation enables us to evaluate GW memory in spacetimes, which do not reduce to AF spacetimes.  

Furthermore, it is easy to verify that the above results match with Bondi shear for Minkowski spacetime~\cite{Flanagan:2015}. $\overline{\phi} (= 2/r)$ is the only nonzero LRS scalar. Substituting $\phi$ in Eq.~(\ref{eq:master-main}) and using the fact that $\dot{N}_{ab}=2\Sigma_{ab}$, we have:
\begin{align} \label{eq:master_Mink}
2 \, \dot{\Sigma}_{ab}  = r\, \G_{ab}~~\Longrightarrow~~ \ddot{N}_{ab} = r \, \G_{ab} \, .
 \end{align}
Considering one polarization ($+$) of the primary (incoming) GW, we set:
\begin{equation}
\label{eq:PrimaryGWProfile}
   {\G}_{\vartheta\vartheta} \equiv \G_{+}=r \cos\vartheta\, \cos(2\varphi) \ddot{h}_I \, ,
\end{equation}
where ${h}_I$ is related to the amplitude of the transient GW event at Earth~\cite{Maggiore:2018sht}. Substituting this in Eq.~\eqref{eq:master_Mink}, we have:
\begin{align}\label{eq:master_Mink_2}
\ddot{N}_{\vartheta\vartheta} (x_-) \equiv \ddot{N}_{+} (x_-) =  r\, f(r,\vartheta,\phi)\,\ddot{h}_{+}^I(x_-) \, .
\end{align}
where, $f(r,\vartheta,\phi)=r\, f(\vartheta,\phi) =r\, \cos\vartheta \cos (2\varphi)$. Integrating the above equation from $t_0$ to $t_1$, we get  $\Delta N_{+} \sim r^2 \Delta h^I$. Since the GW burst falls as $\mathcal{O}(1/r)$, we find the memory observable in our formalism scale as $\mathcal{O}(r)$. Thus, it is a \emph{subleading term}, much akin to the Bondi shear~\cite{Flanagan:2015}. Thus, our analysis reveals the conventional non-oscillatory behavior of the memory observable such that the final spacetime is a shifted Minkowski~\cite{2010-Favata-CQG, Flanagan:2015} (See Fig.`\ref{fig:enter-label} in appendix~\eqref{app:DOF}). We can identify $N_{ab}$ with GW memory in cosmology since ME leads to similar results to traditional methods in Minkowski. We now apply this to cosmology.

\section{Cosmological GW memory}
\label{sec:GW-memory-FLRW}
Since only comoving observers exist in cosmology, the GW and the 2-D surface are comoving. As depicted in Fig.~(\ref{fig:FLRW_mem_model}), an astrophysical process results in the generation of primary GWs at $t = t_0$ (corresponding to redshift $z = z_0$). Assuming that the GWs travel from left to right, the generated GWs perturb the orthogonal 2-D surface. We compute the 2-space shear $\Sigma_{ab}$ at a later time, $t' > t_0$ (corresponding to the redshift $z' < z_0$). We then obtain $N_{ab}$ by integrating 2-D shear tensor $\Sigma_{ab}$ in the range $0 < z < z_0$.

We consider spatially flat FLRW line-element
$ds^2 = 
a^2(\eta)[- d\eta^2 + dr^2 + r^2 d\Omega_2^2]$, where $\eta$ is conformal time, $a(\eta)$ is the scale factor and $d\Omega_2^2$ is the 
unit 2-sphere. Note that $N_{ab}\, dx^a\, dx^b \equiv r^2 d\Omega_2^2$. In FLRW spacetime, $\overline{\Theta} = 3 \H/a, \H= {a^\prime}{/a} \equiv d(\ln a)/d\eta$, and $\overline{\Sigma}= \overline{\phi}=0$. Substituting these in Eq.~\eqref{eq:master-main} reduces to:
\begin{align} \label{eq:master_FLRW}
  2\,  \overline{\Theta} \,\bigg[
   \dot{\Sigma}_{ab} + \dfrac{2}{3} \,   \overline{\Theta}\,\Sigma_{ab} \bigg] - \,(\overline{\mu}+ \overline{p}) \,\Sigma_{ab} =  \G_{ab} \, , 
\end{align}
where dot refers to the retarded time $\tilde{x}_-=\eta-r$ and 
$\overline{\mu}, \overline{p}$ refer to the energy density and pressure of the cosmological fluid. This is the second key expression of this work, regarding which we want to discuss the following points. 
First, the above equations resemble tensor perturbations in FLRW~\cite{1997-Dunsby.Bassett.Ellis-CQG}, except for the source term in the RHS. In cosmology, the tensor perturbations do not have any source and freely propagate. With GW memory, however, we have a primary GW source that produces observable effects in audio frequency.  

Second, unlike Minkowski, $\overline{\phi}$ contributes to the subleading terms. Due to the homogeneity and isotropic nature of the FLRW, $\overline{\Theta}, \overline{\mu}$, and $\overline{p}$ do not have any radial dependence. Hence, the asymptotic (large $r$) behavior of the GW memory in FLRW is different from Minkowski, as the LRS scalars governing their evolution have different fall-offs.
{Lastly, since FLRW only admits comoving observers as mentioned earlier, we do not need to transform the above expression to Eulerian observers~\cite{2023-Jana.Shanki-PRD}.} This was missed in the earlier works on memory in cosmological settings~\cite{Chu:2016,Tolish_Wald_cosmology:2016,Bieri_cosmology:2017,2022-Jokela.Sarkkinen-PRD}. As mentioned earlier, primary GWs generate successive GWs. Since the 2-D surface in FLRW is comoving, the amplitude of the successive waves at each surface will be non-zero. Although the successive waves are weaker, as they travel cosmological distances unabated, their integrated effect is significant.

To quantify this, we consider sharply localized sources \eqref{eq:PrimaryGWProfile} generating primary GWs. Since we are interested in the time evolution of $\Sigma_{ab}$, keeping $r$ fixed, rewriting $\tilde{u}$ in-terms of $\eta$, Eq. \eqref{eq:master_FLRW} leads to:
\begin{equation} \label{eq:shear_FLRW}
   \Sigma_{\vartheta\vartheta}^\prime \, + \Gamma(\eta) \, \Sigma_{\vartheta\vartheta} = ({a(\eta)}/{6\H})\,f(r, \vartheta, \varphi) \, h_{\vartheta\vartheta}^{I^{\prime\prime}}(\eta)
\end{equation}
where, $\Gamma(\eta)=  \dfrac{1}{3}\bigg( \dfrac{{\H^\prime}}{\H}-\H\bigg)$. It is well-known that a field generated by a massless source (like GW) can be modeled as a shock wave satisfying $T_{ab}\propto \delta(\eta-\eta_0)$~\cite{Aichelburg:1970dh,Dray:1984ha,Kocsis:2012}. Here again, $h_{\vartheta\vartheta}^{I^{\prime\prime}}(\eta)$ corresponds to the amplitude of the transient GW event at the 2-D surface. For $+$ polarization, we take $ h_{\vartheta\vartheta}^{I^{\prime\prime}}(\eta) \equiv h_{+}^{\prime\prime} (\eta)= \A_p\, \delta({\eta}-\eta_0)$ where $\A_p$ contains the intrinsic source parameters of the primary GW event~\cite{Maggiore:2018sht}. Substituting this in Eq.~(\ref{eq:shear_FLRW}) and using the two Friedmann equations, we get:
\begin{equation}
    \Sigma_+(\eta) = \frac{\A_p \,f(r,\vartheta,\varphi)}{ \,G(\eta)}\, \rho(\eta_0)
\end{equation}
where, 
$G(\eta)= \exp[{\int_\eta \, \Gamma(\tilde{\eta}) d\tilde{\eta} }],$ 
$\rho(\eta)= G(\eta) a(\eta)/[6\H(\eta)]$. For the comoving observer, using the Lie derivative relation, we have, $\partial_\eta\, ({N_+}/{a^2})=2\Sigma_+/a$.
Defining $\mathcal{N_+}\equiv N_+/(r^2 a^2)$, we find 
\begin{align}
  \mathcal{N_+} =\frac{\A_p f(\vartheta,\varphi) \rho(\eta_0)}{r} \int_{\eta_0}^{\eta_i} 
  \frac{d\eta}{a(\eta) G(\eta)} \, . 
\end{align} 
Rewriting the above expression in terms of the redshift, the GW memory detectable at Earth is:
 \begin{eqnarray}\label{eq:int-memory}
\mathcal{N_+}  =    \frac{h_\oplus}{3E^{2/3}(z_0)} \,\,  \int_0^{z_0} \, dz' \, \frac{(1+z')}{E(z')^{4/3}}.
\end{eqnarray}
where $h_\oplus=\A_p  \, f(\vartheta,\varphi)(1+z)/( \, D_L\,H_0^2) $,  $D_L$ is the luminosity distance [$r=D_L/(1+z)$]~\cite{LIGOScientific:2016wyt}, $E(z)$ is the dimensionless Hubble parameter~\cite{Liu:2024-H0,Croton:2013-H0} and depends on the background cosmology. Eq.~(\ref{eq:int-memory})  gives the GW memory in FLRW spacetimes and is the third key result of this work. 
 This analytical expression allows us to disentangle the GW memory into three distinct components: the GW source amplitude detected at Earth, the cosmological background, and the integrated memory term (from the source redshift to the present). The integrated memory term is analogous to the integrated Sachs-Wolfe effect, where CMB photons gain energy as they pass through evolving large-scale structures in an expanding universe. In both cases, there is an accumulation of effects over cosmological distances due to gravitational clustering. 

\begin{figure}[h]
    \centering
\includegraphics[width=0.8\linewidth]{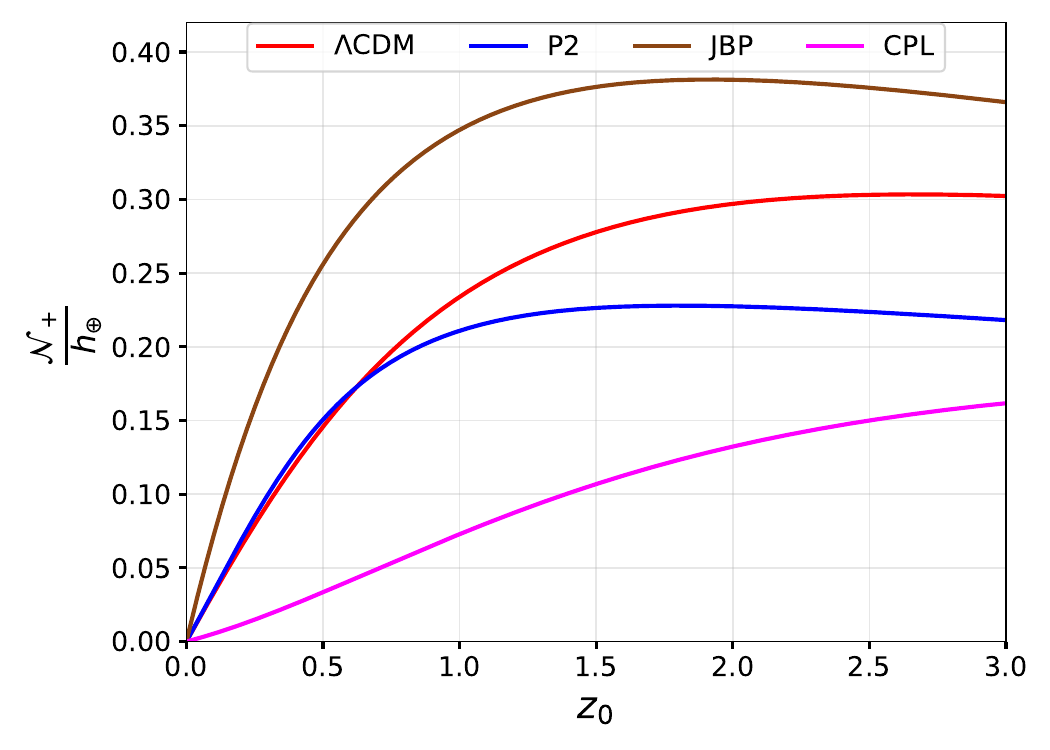}
    \caption{Integrated memory signal for different cosmological models. In the plot $\Omega_m=0.3$ is considered. The standard $\Lambda$CDM  is the limit $w=-1$. For the P2 model,  $w(z)=-1+\frac{(1+z)[\Omega_1+2\Omega_2(1+z)]}{3[\Omega_2 z^2+(\Omega_1+2\Omega_2)z+0.7]}$. The JBP model has $w(z)=w_0+w_1[z/(1+z)^2]$. In CPL model the equation of state becomes, $w(z)=w_0+w_1[z/(1+z)]$. The values of the parameters are: $\Omega_1=-4.162, \Omega_2=1.674$; $w_0=-1.371,w_1=1.127$ (JBP) \cite{Liu:2008-MNRAS}; $w_0=-1.323,w_1=0.745$ (CPL) \cite{Bargiacchi:2021-QSO-BAO:CPL} and $\Omega_m=0.3$.}
    \label{fig:Integrated-memory}
\end{figure} 

{This leads us to the subsequent question: What is the observational implication of Eq.~(\ref{eq:int-memory})? {To answer that, let us  consider a generic cosmological model given by (see, for instance, Ref.~\cite{Liu:2008-MNRAS,Bargiacchi:2021-QSO-BAO:CPL}): 
\begin{align}    
E(z)=   [\Omega_m (1+z)^3+(1-\Omega_m) f(z)]
\end{align}
${\rm  where,~~} f(z)=\exp \left[\int_0^z\, dz'\, \frac{1+w(z')}{(1+z')} \right] \,$.
For a generic cosmological model, the equation of state parameter $w(z)$ evolves as a function of redshift. While there is compelling evidence for the accelerated expansion of the universe, the underlying mechanism driving this acceleration --- the true nature of dark energy --- remains one of the deepest mysteries in cosmology~\cite{Padmanabhan:2002ji,Peebles:2002gy,Sahni:1999}. The $\Lambda$CDM model, although highly successful in resolving numerous cosmological puzzles, has come under growing scrutiny due to increasing tensions between various observational datasets~\cite{DiValentino:2021izs,Kamionkowski:2022pkx}. A natural extension of $\Lambda$CDM involves allowing the equation of state parameter $w(z)$ to vary with redshift. Among the numerous possibilities, we consider three widely studied parametric models where the dark energy EoS evolves with redshift: the Chevallier-Polarski-Linder (CPL) model~\cite{Chevallier:2000-CPL1,Linder:2002-CPL2}, the 2-Parameter (P2) model~\cite{Sahni:2002-P2}, and the Jassal-Bagla-Padmanabhan (JBP) model~\cite{Jassal:2004-JBP}. 

In Fig.~(\ref{fig:Integrated-memory}), we plot ${\cal N}_+/h_\oplus$ as a function of redshift $z$ for these three parametrizations alongside $\Lambda$CDM. From the figure, we observe the following notable features: First, for all parameterizations and $\Lambda$CDM, the integrated memory signal increases monotonically at first with redshift and then saturates around $z=3$. 
For the current value of \(\Omega_m = 0.3\) and a GW signal originating from the same redshift, the integrated GW memory differs across the three parameterizations schemes and the \(\Lambda\)CDM model. Thus, our analysis shows that GW memory emerges as an alternate tool to distinguish between cosmological models and gain deeper insights into the nature of dark energy.} }

Several key distinctions set the present work apart from previous studies on cosmological memory effects. 
Earlier works~\cite{Tolish_Wald_cosmology:2016,Chu:2016,Bieri_cosmology:2017,2022-Jokela.Sarkkinen-PRD}, solved GW perturbation equation for FLRW background. In contrast, our approach derives a master equation that governs the evolution of the transverse 2-space geometry. Furthermore, while those studies primarily consider static observers, our semi-tetrad formalism accounts for comoving observers. This allows us to capture the integrated signal immediately after the onset of the gravitational wave event, a feature that earlier works did not address.  Thus, the total GW amplitude is
    \begin{equation} \label{eq:enhancement}
    h_T= h_\oplus \, (1+\mathcal{N_+}/h_\oplus).
\end{equation}
From the plot in Fig.~(\ref{fig:varying-B}) we infer that the contribution from the memory can be significant for high-redshift objects. We have also included more details in the appendix~\eqref{app:GWmemory-FLRW} describing how our results are different compared to previous predictions.

%
\section{Observational Implications}
\label{Obs-Implications}
 We now ask: How can GW memory be used to constrain cosmological models? To answer this question, we note that as the source redshift approaches $z_0 \to 0$, the integrated memory term and the GW memory signal vanish. At higher redshifts, however, the integrated memory term becomes more prominent. Let us consider two similar BNS mergers: one occurred at $z = 0.01$ (like GW170817) and another at $z = 3$. Assuming a $\Lambda$CDM cosmology, for a source redshift of $z_s = 0.01$, 
$\mathcal{N_+}/h_{\oplus} = 0.0033$. However, for $z = 3$, $\mathcal{N_+}/h_{\oplus}  = 0.302$. This implies the memory signal amplifies around 90 for the $z = 3$ event compared to the $z = 0.01$ event. 
In the case of CPL, JBP and P2 models, the memory signal amplifies around 328, 47 and 65, respectively. Although the ratio $\mathcal{N_+}/h_\oplus$ is smaller than the standard Hubble dilution at lower redshifts, it gains prominence at relatively higher redshifts and hence cannot be ignored. Therefore, GW memory can significantly contribute to detecting high-redshift objects by CE and ET.
 
Secondly, the importance of the integrated memory term becomes evident in multimessenger BNS events~\cite{2019-Nakar-PhysRep,2021-Kalogera-gr-qc}. For BNS systems, using an oscillatory signal-to-noise ratio (SNR) threshold with a lower cutoff of 100, CE and ET are projected to detect $\sim 100$ events annually~\cite{Branchesi:2023,Evans:2023-CE}. Although not every event will have an EM counterpart, a 6-hour observing run with Subaru-HyperSuprimeCam (HSC) is expected to yield a $66\%$ 
 probability of identifying a kilonova counterpart within a sky area of about $100 {\rm deg}^2$\cite{Chan:2015}. For a given cosmological model, EM counterparts can provide the redshift of the source. By stacking multiple high-SNR events and incorporating the integrated memory signal into existing waveform templates, we may establish a promising route to detecting this effect~\cite{Lasky:2016-PRL}. Cross-correlating the redshifts obtained from EM counterparts with those derived from the GW memory signal across various cosmological models could help tighten constraints on the parameter space of different cosmological models. This line of research is particularly timely, given recent DESI results hinting at slight modifications to the standard $\Lambda$CDM paradigm~\cite{DESI:2024_I,DESI:2024l-II,2021-Palmese-APJ}. Additionally, existing waveform models are well known to be susceptible to systematic errors~\cite{Gupta:2024}. While these systematics can be improved with better modeling and numerical simulations,
the integrating memory has yet to be accounted for in these templates. Incorporating this will lead to precise determination of the source redshift. Furthermore, 
recent observations from the James Webb Space Telescope have provided redshift measurements of high-redshift galaxies \cite{Robertson:2022,Gupta:2023-JWST}, enhancing the potential of this method.

Lastly, our approach can be seamlessly applied to \emph{dark sirens}~\cite{Abbott:DarkSiren-2021,DESI:2023,Alfradique:2023,2020-Borhanian-APJL}. With the development of next-generation GW detectors, the localization of GW sources is anticipated to improve substantially. This improved accuracy will facilitate the statistical extraction of redshift information directly from galaxy catalogs, removing the need for an EM counterpart. Hence, comparative analyses of the bright and dark siren methods \cite{Matos:2024,Jin:2023}, especially when incorporating integrated memory signals, offer powerful tools to test the standard $\Lambda$CDM paradigm and explore potential deviations. 

\section{Conclusions}
\label{sec:Conclusions}
Though GW memory is yet to be observed~\cite{Boersma:2020}, it represents a promising avenue for uncovering features of cosmology and gravity. Prior to this work, no unified framework existed for systematically studying GW memory across a wide range of spacetimes. By obtaining a master equation for memory in LRS-II spacetimes, we addressed this gap, demonstrating consistency with traditional approaches in Minkowski spacetime and revealing the prominence of integrated GW memory signals at higher redshifts in cosmological settings.

The $\Lambda$CDM model, while remarkably successful in addressing many cosmological puzzles, faces increasing scrutiny due to emerging tensions between different observational datasets. To date, these observations have largely relied on electromagnetic signals, which, while indispensable, are limited in their ability to probe certain aspects of the cosmic evolution.

The next decade holds the promise of a transformative shift with the advent of GW cosmology, offering a complementary and independent approach to studying the universe. In this context as explicitly shown in this work, the GW memory emerges as a crucial observable. Specifically, we have shown that the amplification of the integrated GW memory is enhanced by a factor of 100 for high-redshift sources, which are within the detection range of next-generation GW observatories like Cosmic Explorer and the Einstein Telescope. By examining a range of dark energy scenarios, we demonstrate that GW memory provides a distinctive probe into the nature of cosmic acceleration, offering fresh insights and a potential resolution to persistent cosmological tensions.


\noindent\emph{\underline{Acknowledgement:}} The authors are grateful to R. Goswami, J. P. Johnson, S. Kar, A. Kushwaha, and S. Mandal for their valuable discussions and feedback on the earlier draft. They also thank R. Kashyap for insightful discussions on multimessenger astronomy and for highlighting the reference \cite{Chan:2015}. I.C. and S.J. acknowledge enriching conversations with  A. Ashtekar, D. Nichols, B. S. Sathyaprakash, J. Veitch. S. J thanks K. Yagi, N. Yunes, A. Nitz and J. Khoury for insightful discussions about GW memory. The work is supported by SERB-CRG (RD/0122-SERB000-044). \href{https://www.et-gw.eu/images/ET-Universe.png}{Thanks to Einstein Telescope team for providing the background for Fig. 1.}

\appendix
\section{Semitetrad covariant formalism}
\label{app:CovariantFormalism}
This section briefly recapitulates the semi-tetrad formalism~\cite{Covariant,2003-Clarkson.Barrett-CQG}, which enables us to study Gravitational Wave (GW) memory for LRS-II spacetime. A crucial feature of the semi-tetrad decomposition is its locality, defined on any open set $\mathcal{S}$. Initially, the properties of spacetime are analyzed relative to a real or fictitious observer whose velocity aligns with the tangent of a timelike congruence, splitting the $4-$d spacetime into a timelike direction and a $3-$space. Subsequently, a preferred spatial direction emerges if the spacetime exhibits certain symmetries, such as local rotational symmetry. The spacetime is further decomposed using this preferred spatial congruence. The field equations are then reformulated in terms of the geometric variables associated with these congruences and the curvature tensor of the spacetime (appropriately decomposed using the congruences).

Although this formalism is well-studied in the literature \cite{clarksonlrs}, for completeness, we provide an overview of the same.

\subsection{Semitetrad 1+3 formalism}\label{A1}
Covariant formalism, first proposed in Refs.~\cite{1955-Heckmann-zap, 1955-A.K.Raychaudhuri-PRL}, later were extensively used in relativistic cosmology~\cite{2012-ellis_maartens_maccallum-Book}. In this formalism the $4-$d spacetime is deconstructed w.r.t a fictitious co-moving observer, moving with velocity $u^a = dx^a/d\tau$($\tau$ is the affine parameter), satisfying $u_a u^a = -1$. The spacetime comprises a timelike congruence $\gamma$ and a $3-$d space orthogonal to $u^a$. The $3-$d space is described by the projection tensor $h_{ab}$ that follows:
\begin{equation}
 g_{ab} = - u_{a}\sss u_{b}+ h_{ab},
\end{equation}
iff the 3-space has no twist or vorticity, $h_{ab}$ becomes metric of the $3-$space. The covariant time derivative along the observers' worldlines, denoted by `${\sss\sss^{\cdot}\sss\sss}$' is defined using the vector ${u^{a}}$, as
\begin{equation}\label{dot}
\dot{Z}^{a ... b}{}_{c ... d} = u^{e}\sss\nabla_{e}\sss Z^{a ... b}{}_{c ... d},
\end{equation} 
for any tensor ${Z^{a...b}{}_{c...d}}$. The fully orthogonally projected covariant spatial derivative, denoted by `\sss${D}$\sss' is defined using the spatial projection tensor ${h_{ab}}$, as
\begin{equation}\label{D}
D_{e}\sss Z^{a...b}{}_{c...d} = h^r{}_{e}h^a{}_{f}\sss...\sss h^b{}_{g}\sss h^p{}_{c}\sss...\sss h^q{}_{d}\sss\nabla_{r}\sss Z^{f...g}{}_{p...q},
\end{equation}
The covariant derivative of the 4-velocity vector ${u^{a}}$ is decomposed irreducibly as follows
\begin{eqnarray}
\nabla_{a}\sss u_{b} &=& -u_{a}\sss A_{b} + \frac{1}{3}h_{ab}\sss\Theta + \sigma_{ab} + \epsilon_{abc}\sss \omega^{c},
\end{eqnarray}
where ${A_{b}}$ is the acceleration, ${\Theta}$ is the expansion of ${u_{a}}$, ${\sigma_{ab}}$ is the shear tensor, ${\omega^{a}}$ is the vorticity vector representing rotation and ${\epsilon_{abc}}$ is the effective volume element in the rest space of the comoving observer. The vorticity vector $\omega^q$ is related to vorticity tensor $\omega^{ab}$ as: $\omega^a \equiv (1/2)\, \epsilon^{abc} \, \omega_{bc}$.

Furthermore, the energy-momentum tensor of matter or fields present in the spacetime, decomposed relative to ${u^{a}}$, is given by
\begin{eqnarray} \label{3.Tab}
T_{ab} &=& \mu\sss u_{a}\sss u_{b} + p\sss h_{ab} + q_{a}\sss u_{b} + u_{a}\sss q_{b} + \pi_{ab},
\end{eqnarray}
where ${\mu}$ is the effective energy density, ${p}$ is the isotropic pressure, ${q_{a}}$ is the 3-vector defining the energy-momentum flux and ${\pi_{ab}}$ is the anisotropic stress. The Weyl tensor also is decomposed into electric part $E_{ab}$ and magnetic part $B_{ab}$ as follows,
\begin{eqnarray}
 E_{ab} &=& C_{acbd}\sss u^{c}\sss u^{d} = E_{<ab>}, \label{E}\\
B_{ab} &=& \frac{1}{2} \varepsilon_{ade}\sss C^{de}{}_{bc}\sss u^{c} = B_{<ab>}. \label{H}
\end{eqnarray}
Here, the angle brackets denote orthogonal projections of vectors onto the 3-space as well as the projected, symmetric, and trace-free (PSTF) part of tensors:
\begin{eqnarray} 
\dot{V}_{<a>}& = &h_{a}{}^{b} \sss \dot{V}_{b},\label{angbrac1}\\ 
Z_{<ab>}& =& \bra{h^{c}{}_{(a}\sss h^{d}{}_{b)} - \frac{1}{3}h_{ab}\sss h^{cd}}\sss Z_{cd}.\label{angbrac2}
\end{eqnarray}

\subsection{Semitetrad 1+1+2 formalism}\label{A2}

The $3-$space mentioned in the $1+3$ covariant formalism can be further split into one spacelike direction $e_a$ satisfying $e_a e^a = 1$ and a $2-$d surface orthogonal to both $u^a$ and $e^a$. 
The 1+1+2 covariantly decomposed spacetime is expressed in terms of the projection tensor $N_{ab}$ associated with the $2-$d surface as: 
\begin{equation}\label{2.Nab}
g_{ab} = -u_{a}\sss u_{b} + e_{a}\sss e_{b} + N_{ab},
\end{equation}
where ${N_{ab}}$ ${\left(e^{a}\sss N_{ab} = 0 = u^{a}\sss N_{ab}, N^{a}{}_{a} = 2\right)}$ projects vectors onto the 2-sheets, orthogonal to ${u^{a}}$ and ${e^{a}}$.  We introduce two new derivatives for any tensor ${\phi_{a...b}{}^{c...d}}$:
\begin{eqnarray}
\label{hatderiv}
\hat{\phi}_{a..b}{}^{c..d} &\equiv& e^{f}\sss D_{f}\sss \phi_{a..b}{}^{c..d}, \\
\label{deltaderiv}
\hspace*{-0.6cm}\delta_{f}\phi_{a...b}{}^{c...d} &\equiv& N_{f}{}^{j} N_{a}{}^{l} ... N_{b}{}^{g} N_{h}{}^{c} ...N_{i}{}^{d}  D_{j}\phi_{l...g}{}^{h...i}. \nonumber \\
\end{eqnarray}
Eq.(\ref{hatderiv}) denotes the derivative along the preferred spacelike direction $e^a$ while 
$\delta_f$ in Eq.(\ref{deltaderiv}) gives the 2-space derivative.

The  1+3 geometrical and dynamical quantities and anisotropic fluid variables are split irreducibly as
\begin{eqnarray}
A^{a} &=& \mathcal{A}\sss e^{a} + \mathcal{A}^{a}, \\
\omega^{a} &=& \Omega\sss e^{a} + \Omega^{a}, \\
\sigma_{ab} &=& \Sigma\left(e_{a}\sss e_{b} - \frac{1}{2}\sss N_{ab}\right) + 2\sss\Sigma_{(a}\sss e_{b)} + \Sigma_{ab}, \\
E_{ab} &=& \mathcal{E} \left(e_{a}\sss e_{b} - \frac{1}{2}\sss N_{ab}\right) + 2\sss\mathcal{E}_{(a}\sss e_{b)} + \mathcal{E}_{ab}, \\
B_{ab} &=& \mathcal{H}  \left(e_{a}\sss e_{b} - \frac{1}{2}\sss N_{ab}\right) + 2\sss\mathcal{H}_{(a}\sss e_{b)} + \mathcal{H}_{ab}\,.
\end{eqnarray}
 
The fully projected 3-derivative of ${e^{a}}$ is given by
\begin{eqnarray}
D_{a}\sss e_{b} &=& e_{a}\sss a_{b} + \frac{1}{2}\sss\phi\sss N_{ab} + \xi\sss\varepsilon_{ab} + \zeta_{ab},
\end{eqnarray}
where traveling along ${e^{a}}$, ${a_{a}}$ is the sheet acceleration, ${\phi}$ is the sheet expansion, ${\xi}$ is the vorticity of ${e^{a}}$ (the twisting of the sheet) and ${\zeta_{ab}}$ is the shear of ${e^{a}}$. 

We can immediately see that the Ricci identities and the doubly contracted Bianchi identities, which specify the evolution of the complete system, can now be written as the time evolution, spatial propagation, and spatial constraints of an irreducible set of geometrical variables:
\begin{align}
\label{Dgeom}
\mathcal{D}_{geom} = \{\Theta, \sss \mathcal{A}, \sss\Omega, \sss\Sigma, \sss\mathcal{E}, \sss\mathcal{B}, \sss\phi, \sss\xi, \sss\mathcal{A}_{a}, \sss\Omega_{a}, \sss\Sigma_{a}, 
\sss\alpha_{a}, \sss a_{a}, \sss\mathcal{E}_{a}, \sss\mathcal{B}_{a},  \sss\Sigma_{ab}, \sss\zeta_{ab}, 
\sss\mathcal{E}_{ab}, \sss\mathcal{B}_{ab}\}
\end{align}

\subsection{Degrees of freedom of incoming GW}
\label{app:DOF}

\begin{figure}[h]
    \centering \includegraphics[scale=0.5]{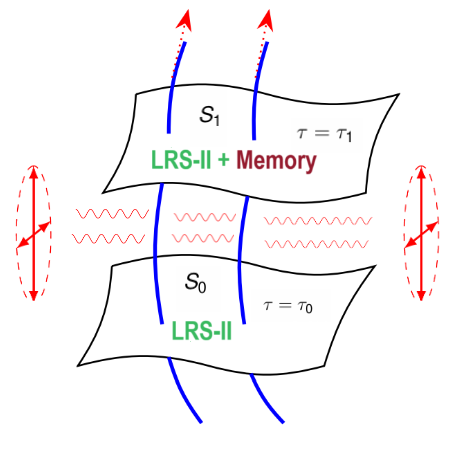}
    \caption{The presence of GW alters the background LRS-II spactime.}
    \label{fig:enter-label}
\end{figure}

Let us consider the following scenario shown in Fig. \eqref{fig:enter-label}. LRS-II spacetime is perturbed due to the incoming GW. We need to know the true GW degrees of freedom (DOF) in this formalism to identify the GW memory. In this section, we identify the true GW DOF for completeness.

Let the initial LRS-II spacetime be represented by the geometric variables $\mathcal{D}_{0}$. Due to the incoming GW, the spacetime is no longer LRS-II. Since the incoming GW is weak, we can assume that the resulting spacetime will perturb LRS-II. Let us represent the geometric variables of this spacetime as $\mathcal{D}_{\rm n}$. From this point we will add `overline' to all the background variables in the LRS-II spacetime. 

The GW mimicking variable may consist of one or several vector or tensor quantities listed in $\mathcal{D}_{\rm n}$. As described in Refs.~\cite{2011-Owen.Thorne-PRL,Thorne2}, the Weyl curvature tensor $C_{abcd}$ characterizes the gravitational distortion or perturbation caused by GWs. 
Both the electric and magnetic parts of the Weyl tensor follow a closed form of the wave equation in the FLRW spacetime~\cite{1966-Hawking-ApJ, 2019-Goswami.Ellis-CQG}.

In the $1+3$ formalism, $H_{ab}$ and $E_{ab}$ represent certain features of GWs, while in the $1+1+2$ formalism, their components such as $\B,\B_a, \B_{ab}$ and $\overline{\E}, \E_a, \E_{ab}$ \footnote{In LRS-II spacetimes, the background LRS scalar $\B=0$. Hence, it not denoted by $\overline{\B}$. The contribution can come from GW perturbation.} carry information about the passing GWs during $[\tau_0,, \tau_1]$. As mentioned earlier, GWs perturb the $2-$dimensional surface $S_{0}$ orthogonal to it. Hence, the vector and tensor components of $H_{ab}$ and $E_{ab}$ in the $2-$dimensional surface ($\B_a, \B_{ab}$) and ($\E_a, \E_{ab}$) describe the GW distortion during the time interval $[\tau_0,, \tau_1]$. 
 
As discussed above, in the $1+1+2$ formalism $\E_{a}, \E_{ab}, \B_a, \B_{ab}$ each have two components, resulting in a total of 8 DOF. However, considering the 2 constraint equations for the scalar part of electric and magnetic Weyl~\eqref{eq:EScalar-delu},\eqref{eq:HScalar-delu}, the DOF reduces to 6. Furthermore, the relationship among the components of the electric and magnetic parts of the Weyl tensor, as given by Eq.\eqref{eq:Evec-Hvec}, \eqref{eq:Eten-Hten} in the Sec. \eqref{app:formulation-masterEq}, reduces the DOF to 2, matching the DOF of GWs in linearized gravity theory~\cite{Thorne1,Thorne2}. In principle, any of $\E_{a}, \E_{ab}, \B_a , \B_{ab}$ can represent the passing gravitational wave. However, to formulate the master equation for GW memory as done in Sec.~\eqref{app:formulation-masterEq}, we choose $\B_a$ to describe the incoming GW.

\section{Relation between GW memory and shear}
\label{app:memory-shear}

GW memory refers to the change in the spacetime metric resulting from the passage of GWs. Mathematically, it is defined as:
\begin{align}
\Delta h_{+, \times}^{\mathrm{mem}}=\lim _{t \rightarrow+\infty} h_{+, \times}(t)-\lim _{t \rightarrow-\infty} h_{+, \times}(t),
\end{align}
where $t$ represents the time of the asymptotic (static) observer, and $h_{+}$ and $h_{\times}$ denote the plus and cross modes of the GW signal, respectively.

However, in covariant formalism, $u^a$ defines the time. Any vector or tensor change with respect to time can be obtained from the Lie derivative of any tensor or vector along $u^{a}$. To quantify gravitational memory and relate it to the geometrical variables described in $\mathcal{D}_{\rm n}$, we begin with the Lie derivative of $N_{ab}$ along $u^{a}$.
\begin{align}
     \mathcal{X}_{ab} \equiv \mathcal{L}_{u} N_{ab} = u^{c}\nabla_{c} N_{ab} + N_{cb}\nabla_{a} u^{c} + N_{ca}\nabla_{b} u^{c}
\end{align}
Substituting the geometrical variables described in $1+1+2$ formalism, we obtain,
\begin{align}
\mathcal{X}_{ab} = 2 n_{(a} \left[-\alpha_{b)} + \Sigma_{b)} - \lc_{b)\,  c} \, \Omega^{c}\right] + 2 \, \Sigma_{ab} 
   + N_{ab} \left[\frac{2 \overline{\Theta}}{3}  - \overline{\Sigma}\right]
\end{align}
The brackets $()$ in the subscript denote the symmetrization of the terms. In Ref.~\cite{clarksonlrs}, the author mentions the existence of a $2-$surface if the Greenberg vector $\Sigma^{a} + \lc^{ab} \Omega_{b} - \alpha^{a} = 0$. Using this Greenberg vector condition in the previous equation, we obtain:
\begin{align}
\mathcal{X}_{ab} = - 4 n_{(a} \lc_{b) c} \, \Omega^{c} + 2 \Sigma_{ab} + N_{ab} \left(\frac{2 \overline{\Theta}}{3} - \overline{\Sigma}\right)
\end{align}
The quantity $\mathcal{X}_{ab}$ above is symmetric but not transverse-traceless. Its transverse trace-less part is derived as follows:
\begin{align}
\mathcal{X}_{\{ab\}} = N^{\alpha}\,_{(a} N^{\beta}\,_{b)} \mathcal{X}_{cd} - \frac{1}{2}N_{ab} N^{cd} \mathcal{X}_{cd}
\end{align}
Substituting $\mathcal{X}_{\alpha\beta}$ into the RHS of the above equation, we finally obtain:
\begin{align}
    \mathcal{X}_{\{ab\}} = 2 \,\Sigma_{ab}\label{eq:shear-to-mem-app}
\end{align}
The $\Sigma_{ab}$ tensor is an indicator of memory in this scenario as it represents the time evolution of the projection tensor onto the $2-$space. We need to establish the conditions under which the projection tensor $N_{ab}$ reduces to the metric of the $2-$space. These conditions are: 
1) the Greenberg vector $\Sigma^{a} + \lc^{ab} \Omega_{b} - \alpha^{a} = 0$, and 2) $\xi= \Omega=0$).

Similarly, the covariant derivative of $N_{ab}$ projected along the spacelike direction will be $\zeta_{ab}$, where $\zeta_{ab}$ represents the shear related to the spacelike direction $n^{a}$. 

In the following calculation, we assume that the projection tensor $N_{ab}$ reduces to the metric of the $2-$space if:
\begin{align}
\label{eq:2d-condition1}
\text{Greenberg vector:} \quad  \Sigma^{a} + \lc^{ab} \Omega_{b} - \alpha^{a} &= 0,
\\
\label{eq:2d-condition2}
\text{} \xi = \Omega &=0 
\end{align}
are satisfied.

\section{Formulation of gravitational memory in \emph{LRS-II} spacetime}
\label{app:formulation-masterEq}

As mentioned earlier, the GW burst perturbs the LRS-II spacetime; hence, the final spacetime is no longer LRS-II.
Next,  we formulate the master equation to obtain GW memory in the \emph{LRS-II} spacetime using the evolution, propagation, and constraint equations of electric $\overline{\E},\, \E_{a},\, \E_{ab}$ and magnetic parts $\B,\, \B_{a},\, \B_{ab}$ of Weyl curvature tensor. Here, we use the equations derived in Ref.~\cite{clarksonlrs} considering terms till first order w.r.t. the quantities listed in $\mathcal{D}_{\rm n}$. We assume that the \emph{LRS-II} background contains only the energy density and pressure in the energy-momentum tensor that is a function of time, measured along $u^a$. We ignore the anisotropic stress and heat flux. The propagation equations for $\overline{\E}, \, \B$ are:
\begin{align}
\hat{\overline{\E}} &= - \delta_{a}\E^{a} - \frac{3}{2} \overline{\phi} \overline{\E} \label{eq:EScalar-hat}\\
\hat{\B} &= - \delta_{a}\B^{a} - \frac{3}{2} \overline{\phi} \B \label{eq:BScalar-hat}
\end{align}
As GWs propagate along the null direction, instead of using the time coordinate $t$ and radial coordinate $r$, we utilize the null coordinate $x_- \equiv t - r$. At the asymptotic limit, the time and radial derivatives can be expressed as $\partial_{t} \sim \partial_-$ and $\partial_{r} \sim -\partial_{-}$, where $\partial_{x_-}\equiv \partial_-$. Using this notation, we rewrite the propagation equations as follows:
\begin{align}
\partial_{-}\overline{\E} - \frac{3}{2} \overline{\phi}~ \overline{\E} &= \delta_{a}\E^{a} \label{eq:EScalar-delu}\\
\partial_{-} \B - \frac{3}{2} \overline{\phi} \B &= \delta_{a}\B^{a} \label{eq:HScalar-delu}
\end{align}
These are the constraint equations relating the scalar part $\overline{\E}(\B)$ to the vector $\E_{a}(\B_{a})$. Now, we aim to establish if there exists any relation between the electric and magnetic parts of the Weyl tensor. To proceed, we derive the evolution equations for $\E_{a}$ and $\B_{a}$ after incorporating conditions (\ref{eq:2d-condition1}) and (\ref{eq:2d-condition2}) in the null coordinate.
\begin{align}
\label{eq:Evec-dot}
\partial_{-} \E_{a} - \frac{1}{2} \lc_{ab} \partial_{-} \B^{b}  = \frac{3}{4} \lc_{ab}  \delta^{b} \B + \frac{1}{2} \lc_{bc} \delta^{b} \B^{c}\,_{a} -\frac{1}{2} (\overline{\mu}+ \overline{p}) \, \Sigma_a  + 
 \left(\frac{3}{4} \overline{\Sigma} - \overline{\Theta}\right) \E_{a} - \left(\frac{1}{4} \overline{\phi} + \overline{\A} \right)\lc_{ab} \B^{b} 
\end{align}
\begin{align}
\label{eq:Hvec-dot}
\partial_{-} \B_{a} + \frac{1}{2} \lc_{ab} \partial_{-} \E^{b} = -\frac{3}{4} \lc_{ab}  \delta^{b} \overline{\E} - \frac{1}{2} \lc_{bc} \delta^{b} \E^{c}\,_{a}  + \left(\frac{3}{4} \overline{\Sigma} - \overline{\Theta}\right) \B_{a} + \left(\frac{1}{4} \overline{\phi} + \overline{\A} \right)\lc_{ab} \E^{b} 
\end{align}

We write the propagation equations of $\E_{a},\, \B_{a}$ in terms of \emph{null coordinate} as follows,
\begin{align}
\label{eq:Evec-hat}
\partial_{-} \E_{a}  &= -\frac{1}{2}\delta_{a}\overline{\E} + \delta^{b}\E_{ab} + \frac{3\overline{\phi}}{2}  \E_{a} + \frac{3\overline{\Sigma}}{2} \, \lc_{ab}\B^{b}
\end{align}
and, 
\begin{align}
\label{eq:Hvec-hat}
\partial_{-} \B_{a} &=-\frac{1}{2}\delta_{a}\B + \delta^{b}\B_{ab} +\frac{3}{2} \overline{\phi} \B_{a} - \frac{3\overline{\Sigma}}{2}  \lc_{ab}\E^{b}
\end{align}
Now, multiplying Eq.~\eqref{eq:Hvec-dot} with $\lc^{ae}$ and adding it to Eq.~\eqref{eq:Evec-dot} (by setting the same dummy index to $a$) we obtain the following relation,
\begin{align}
\label{eq:Evecd-otepsilonHvec-dot}
\frac{3}{2} \partial_{-}\left(\E_{a}-\lc_{ab} \B^b \right) = & \frac{3}{4}\left( \lc_{ab}\delta^b \B - \delta_{a} \overline{\E}\right) -\frac{1}{2}\left(\overline{\mu} + \overline{p} \right)\Sigma_{a}  
- \frac{3}{4}\left(\alpha_{a} -2 \A_{a}\right)\overline{\E} 
 +\nonumber\\
& \left(\frac{\overline{\phi}}{4} + \overline{\A}  + \frac{3}{4} \overline{\Sigma} - \overline{\Theta} \right)\left(\E_a - \lc_{ab} \B^b \right) 
 + \frac{1}{2} \left(\lc_{bc}\delta^b \B^{c}\,_{a} - \delta^b \E_{ab}\right)
\end{align}
Substituting Eqs.~\eqref{eq:Evec-hat} and \eqref{eq:Hvec-hat} into the above equation and multiplying the result by $\lc^{ad}$ we obtain,
\begin{align}
\label{eq:E-H-relation}
2 \delta^{b} \left(\E_{ab} - \lc^{ad} \B^{d}\, _{b}\right) + 2 (\overline{\mu} + \overline{p}) \Sigma_{a}  = \left(-2\overline{\phi} + \overline{\A} + 3\overline{\Sigma} - \overline{\Theta} \right)\left(\E_a - \lc_{ab} \B^b \right)
\end{align}
  One can check that $\alpha_a$=0 for perturbed FLRW and Minkowski spacetimes. This automatically ensures that $\Sigma_a=0$ if  $\Omega_a=0$ from the Greenberg vector condition. We choose in the rest of our work the following relation between $\{\E_a, \, \B_a\}$ and $\{\E_{ab}, \, \B_{ab}\}$.
\begin{align}
\label{eq:Evec-Hvec}
\E_a = \lc_{ab} \B^b\\
\label{eq:Eten-Hten}
\E_{ab} = \lc_{ad} \B^{d}\,_{b}
\end{align}
We have the evolution equation of $\E_{ab}$, tensor part of the electric part of Weyl tensor in \emph{null coordinate} as,
\begin{align} 
& \partial_- \E_{\lb ab\rb} + \partial_- \left(\lc_{c\lb a}\B_{b\rb}^{~~c}\right) = -\lc_{c\lb a}\delta^{c}\B_{b\rb} -\f32 \overline{\E}\Sigma_{ab}  \nonumber \\ 
&- \f32 \B \lc_{c \lb a} \zeta_{b\rb}^{~~c} - \bra{\overline{\Theta} + \f32 \overline{\Sigma}} \E_{ab} 
 +\bra{\f12 \overline{\phi}+2 \overline{\udota}}\lc_{c\lb a}\B_{b\rb}^{~~c} -\f12 (\overline{\mu} + \overline{p}) \Sigma_{ab} \label{Edot} 
\end{align}
We have assumed $\partial_- \lc_{ab} =0$. Now substituting Eq.~\eqref{eq:Eten-Hten} relation and simplifying the above equation we obtain,
\begin{align}
\label{eq:master1}
3 \overline{\E} \, \Sigma_{ab} + \left(2\overline{\Theta} + 3 \overline{\Sigma} + \overline{\phi} + 4 \overline{\A} \right)\E_{ab} + \left(\overline{\mu} + \overline{p}\right)\Sigma_{ab} 
= -2 \lc_{c\lb a} \delta^c \, \B_{b\rb}
\end{align}
We already mentioned that the shear tensor $\Sigma_{ab}$ holds information about the spacetime alteration or the memory of the passing GW. Hence we want to re-write the above Eq.~\eqref{eq:master1} substituting $\E_{ab}$ in terms of $\Sigma_{ab}$. To do so, we have the evolution equation of $\Sigma_{ab}$ as, 
  \begin{align}
\label{eq:Sigma-dot-app}
\dot\Sigma_{\lb ab \rb} = 
\delta_{\lb a}\udota_{b\rb}
-\left(\f23 \overline{\Theta} + \f12 \overline{\Sigma}\right)\Sigma_{ab} + \overline{\A} \zeta_{ab}-\E_{ab}
\end{align}  

We set $\A= \A_a =0$ in the above equation and obtain, 
\begin{align}
\label{eq:E-Sigma-relation}
\E_{ab} = -\dot\Sigma_{\lb ab \rb}
-\left(\f23 \overline{\Theta} + \f12 \overline{\Sigma}\right)\Sigma_{ab} 
\end{align}
Substituting Eq.~\eqref{eq:E-Sigma-relation} into Eq.~\eqref{eq:master1} we obtain the final master equation for GW memory, 
\begin{align}
\label{eq:master2}
 3 \overline{\E} \, \Sigma_{ab} - \left(2 \overline{\Theta} + 3 \overline{\Sigma} + \overline{\phi} \right)\dot\Sigma_{\lb ab \rb}+ \left(\mu + p\right)\Sigma_{ab}
 -\left(2\overline{\Theta} + 3 \overline{\Sigma} + \overline{\phi}  \right) \left(\f23 \overline{\Theta} + \f12 \overline{\Sigma}\right) \Sigma_{ab} = -2 \lc_{c\lb a} \delta^c \, \B_{b\rb}
\end{align}

\section{Cosmological GW memory signal}
\label{app:GWmemory-FLRW}
\subsection{Details of the GW memory equation in FLRW spacetime}

The master equation in FLRW is
\begin{equation} \label{eq:master-rough}
    2 \overline{\Theta} \bigg[\dot{\Sigma}_{ab} + \frac{2}{3}\overline{\Theta}\, \Sigma_{ab}\bigg] - (\overline{\mu} + \overline{p}) \, \Sigma_{ab} = 2 \, \G_{ab}
\end{equation}
where $\G_{ab} = 2 \epsilon_{c\{a} \delta^c \B_{b\}}$ and the dot is basically derivative w.r.t. $\nabla$.
$$\dot{\Sigma}_{ab}= u^c\, \nabla_c \,\Sigma_{ab} = \frac{1}{a} \bigg(\partial_\eta \Sigma_+- \frac{2a^\prime}{a} \Sigma_+\bigg)$$
Here, "+" denote $\vartheta\vartheta$ component. 
In $c=8\pi G=1$ units, the Friedmann equations are: 
$$\overline{\mu} + \overline{p} =2\, \bigg[ \bigg(\dfrac{2{a^\prime}^2}{a^4}\bigg)-\bigg(\dfrac{a^{\prime\prime}}{a^3}\bigg)\bigg]$$

Finally, in conformal time coordinate, Eq.(\ref{eq:master-rough}) becomes,
\begin{equation}
\label{eq:master-rough3}
    \partial_\eta\, \Sigma_+ \, +  \Gamma(\eta) \Sigma_+\, = \frac{a(\eta)}{6\H} r \cos\vartheta\, \cos(2\varphi)\, h_I^{\prime\prime}
\end{equation}
where
\begin{equation}
\label{def:Gamma}
\Gamma(\eta) = \bigg[ 
  \frac{1}{3} \bigg(\frac{\H^\prime}{\H} -\H\bigg]\bigg) \,    
\end{equation}  

Let's define, $G(\eta) \equiv  \text{Exp}[\int_\eta \, \Gamma(\tilde{\eta})\, d\tilde{\eta} ]$.
Multiplying $\Gamma(\eta)$ on both sides of Eq. \eqref{eq:master-rough3}, we get, 
\begin{align}
    \partial_\eta\,[\Sigma_+ G(\eta)]= f(r,\vartheta,\varphi) \bigg[G(\eta)\frac{a(\eta)}{6\H}\bigg]\, h_I^{\prime\prime}
\end{align}
Let $\rho(\eta)=\bigg[G(\eta)\frac{a(\eta)}{6\H(\eta)}\bigg]$ and $f(r,\vartheta,\varphi)=r f(\vartheta,\varphi)= r\, \cos(\vartheta)\, \cos(2\varphi)$. The final solution,
\begin{align}
\label{eq:sigma-prime}
  \Sigma_+(\eta)= \frac{f(r,\vartheta,\varphi)}{G(\eta)} \,\int\,d\tilde{\eta} \, \rho(\tilde{\eta})  \, h_I^{\prime\prime}(\tilde{\eta})
\end{align}
Now, for shock wave pulses, $T_{ab}\sim\, \delta(\eta-\eta_0)$. Thus, the Ricci tensor also scales as $\delta(\eta-\eta_0)$. We know that, $(\text{Riemann})\sim h_I^{\prime\prime}\sim \delta(\eta-\eta_0)$. Thus, we take $ h_I^{\prime\prime} (\eta)=\A_p\, \delta({\eta}-\eta_0)$. Thus Eq.~\eqref{eq:sigma-prime} becomes,
\begin{equation}
    \Sigma_+(\eta) = \frac{\A_p \,f(r,\vartheta,\varphi)}{G(\eta)}\, \rho(\eta_0)
\end{equation}
Using the relation $\partial_\eta\, ({N_+}/{a^2(\eta)})=2\Sigma_+/a(\eta)$ we obtain the following,
\begin{align}
 & \frac{N_+}{a^2(\eta)} = \A_p f(r,\vartheta,\varphi) \, \rho(\eta_0) \int_{\eta_0}^{\eta_i} \frac{d\eta}{a(\eta)\, G(\eta)}  \nonumber 
 \\
 &  =\A_p f(r,\vartheta,\varphi) \, \rho(\eta_0) \int_{\eta_0}^{\eta_i} \frac{d\eta}{a(\eta)}\, \text{Exp}\bigg[-\int^\eta d\tilde{\eta}\, \Gamma(\tilde{\eta})\bigg]
\end{align}

\subsection{FLRW memory in terms of redshift}

Setting the current scale factor to unity, we know:
\begin{equation}
a(t) = \frac{1}{1+z}, \quad  \Longrightarrow \quad da(t) = - \frac{dz}{(1+z)^{2}}   
\end{equation}
Rewriting $dt$ as
\begin{equation}
dt = \frac{dt}{da} da(t)  
\end{equation}
we get:
\begin{equation}
dt = -  \frac{1}{(1+z)^{2}} \frac{dz}{H} \frac{1}{a(t)} 
= -  \frac{1}{(1+z)} \frac{dz}{H} 
\end{equation}
%
Thus, the cumulative memory tensor in terms of redshift is given by 
\begin{eqnarray}
\mathcal{N_+}  = \frac{N_+}{a^2 r^2} =  \frac{h_\oplus}{3E^{2/3}(z_0)} \,\,  \int_0^{z_0} \, dz' \, \frac{(1+z')}{E(z')^{4/3}}.
\end{eqnarray}
where $h_\oplus=\A_p  \, f(\vartheta,\varphi)(1+z)/( \, D_L\,H_0^2) $,  $D_L$ is the luminosity distance [$r=D_L/(1+z)$] and $H(z)=H_0 E(z)$ with $E(z)=\sqrt{\Omega_m(1+z)^{2/B}+ \Omega_\Lambda}$.

\subsection{Memory signal versus source redshift for a minimal cosmological model}

\begin{figure}
    \centering
    \includegraphics[width=0.7\linewidth]{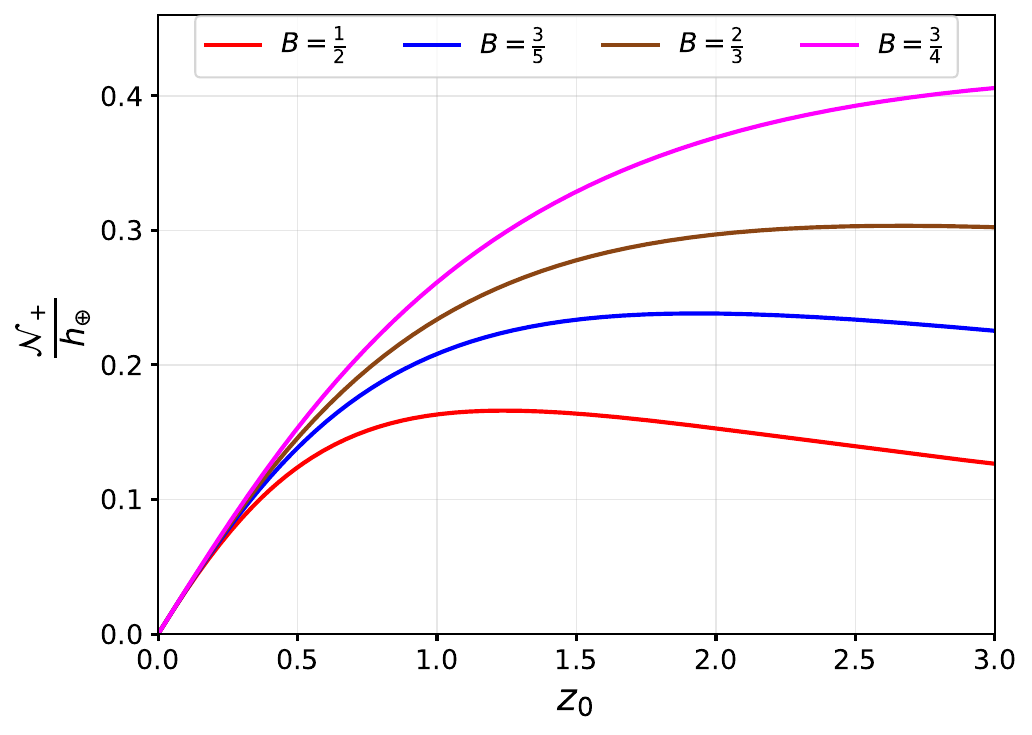}
    \caption{Integrated memory signal by varying the parameter $B$.}
    \label{fig:varying-B}
\end{figure}

In Fig.~\ref{fig:Integrated-memory} of the sec.~\eqref{sec:GW-memory-FLRW}, we have shown how the integrated memory signal varies for different cosmological models  that are explored to constrain the equation of state of dark energy using different cosmological data. In this Appendix, we  focus on a comparatively simplified, yet generic cosmological model~\cite{Sen:2001-PLB,Mukhopadhayay:2024}, 
described by the expansion rate   
\[
E(z)=   [\Omega_m (1+z)^{2/B}+(1-\Omega_m)] . 
\]
In this parametrization, $B=2/3$ gives $\Lambda$CDM. The integrated memory signal for different values of $B$ is shown in Fig.~\ref{fig:varying-B}. The figure illustrates that the signal increases with increasing $B$. For comparison, if the universe were dominated by radiation and dark energy, the signal would be weaker than that of the current universe, which is predominantly composed of matter and dark energy.

\subsection{Comparison of the results with earlier approaches}
\label{app:Comparison}

The GW memory $\mathcal{N_+}$ obtained in the sec.~\eqref{sec:GW-memory-FLRW} is an integrated effect. This manifests as a correction over the primary standard GW signal. Moreover, also note that the $h_\oplus$ term captures the standard Hubble dilution in cosmology as,
\begin{equation} \label{eq:hubble-dilution}
     h_\oplus = (1+z) \frac{\mathcal{A} f(\vartheta,\varphi)}{3\, D_L H_0^2} 
\end{equation}
where $D_L$ is the luminosity distance. Thus, including the memory, the  GW amplitude ($h_T$) will be,
\begin{equation}
    h_T= h_\oplus \, \bigg(1+\frac{\mathcal{N_+}}{h_\oplus} \bigg).
\end{equation}
Although this correction is small compared to the signal, it remains significant and cannot be ignored, particularly in the context of probing the high-redshift universe, as demonstrated in Fig.~\ref{fig:varying-B}. 

Additionally, our approach presents several key distinctions from earlier works, including those in Refs. \cite{2022-Jokela.Sarkkinen-PRD, Tolish_Wald_cosmology:2016, Bieri_cosmology:2017, Chu:2016, Kehagias:2016}. First, while these prior studies solve the GW perturbation equation within the framework of an FLRW background, our method derives a master equation that governs the evolution of the transverse $2-$space. Second, earlier approaches focus on static observers, whereas our semi-tetrad formalism considers comoving observers associated with the fundamental conformal Killing vector field, $U^a$, in FLRW spacetime. This refinement enables us to capture the integrated signal immediately following the gravitational wave event feature absent in previous treatments. Third, while the formalism in Ref. \cite{Chu:2016} allows for extensions to higher-dimensional settings, our formalism is specifically designed for 4-dimensional LRS spacetimes, without the same generalizability. Lastly, our approach offers flexibility, making it readily applicable to other cosmological spacetimes beyond FLRW and to different cosmological models.

\input{arxiv-erratum-corrected.bbl}
\end{document}

%% file: arxiv-erratum-corrected.bbl
%